\documentclass[conference]{IEEEtran}
\IEEEoverridecommandlockouts
\usepackage{cite}
\usepackage{amsmath,amssymb,amsfonts}
\usepackage{algorithmic}
\usepackage{graphicx}
\usepackage{textcomp}
\usepackage{xcolor}
\def\BibTeX{{\rm B\kern-.05em{\sc i\kern-.025em b}\kern-.08em
    T\kern-.1667em\lower.7ex\hbox{E}\kern-.125emX}}
\begin{document}

\title{A cost effective eye movement tracker based wheel chair control algorithm for people with paraplegia\\
}
\author{\IEEEauthorblockN{Skanda Upadhyaya}
\IEEEauthorblockA{\textit{Department of Electrical and Electronics Engineering} \\
\textit{National Institute of Technology Karnataka}\\
Surathkal, India }\\
\and
\IEEEauthorblockN{Siddhanth P. Rao}
\IEEEauthorblockA{\textit{Department of Electrical and Electronics Engineering} \\
\textit{National Institute of Technology Karnataka}\\
Surathkal, India }\\
\and
\IEEEauthorblockN{Shravan Bhat}
\IEEEauthorblockA{\textit{Department of Electrical and Electronics Engineering} \\
\textit{National Institute of Technology Karnataka}\\
Surathkal, India }\\
\and
\IEEEauthorblockN{V Ashwin}
\IEEEauthorblockA{\textit{Department of Electrical and Electronics Engineering} \\
\textit{National Institute of Technology Karnataka}\\
Surathkal, India }\\
\and
\IEEEauthorblockN{Krishnan Chemmangat}
\textit{Assistant Professor}
\IEEEauthorblockA{\textit{Department of Electrical and Electronics Engineering} \\
\textit{National Institute of Technology Karnataka}\\
Surathkal, India \\
cmckrishnan@nitk.edu.in}
}

\maketitle

\begin{abstract}
Spinal cord injuries can often lead to quadriplegia in patients limiting their mobility. Wheelchairs could be a good proposition for patients, but most of them operate either manually or with the help of electric motors operated with a joystick. This, however, requires the use of hands, making it unsuitable for quadriplegic patients. Controlling the eye movement, on the other hand, is retained even by the people who undergo brain injury. Monitoring the movements in the eye can be a helpful tool in generating control signals for the wheelchair. This paper is an approach on converting obtained signals from the eye into meaningful signals by trying to control a bot which imitates a wheelchair. The overall system is cost-effective and uses simple image processing and pattern recognition to control the bot. An android application is developed, which could be used by the patients' aid for more refined control of the wheelchair in the actual scenario.
\end{abstract}


\section{Introduction}
{\Huge\textbf A}ccording to a study conducted by American Spinal Injury Association in 2016 it was found that a total of 12500 people are affected by the spinal cord injuries in varying degree of spinal cord injuries \cite{roberts2017classifications}. According to their estimates, the global yearly average is anywhere between 1.3 to 2.6 hundred thousand cases. Many of these injuries lead to quadriplegia which severely affects the mobility of a person. Apart from trauma, quadriplegia can be inherited \cite{blackstone2018hereditary}. A motorised wheelchair appropriately controlled by signal generated by the patients body could be an effective solution for this problem. Control signals such as electromyogram (EMG) \cite{jang2016emg}, electroencephalogram (EEG) \cite{al2018review},  or electrooculogram (EOG) \cite{huang2017eog}, generated from the quadriplegic patients eyes, face or head could be used to control this device. However, capturing and processing minute signals is often difficult and the sensing and signal conditioning circuits are highly sophisticated and expensive making the affordability a big concern. In a country like India where the total number of physically disabled people was found to be close to 5.36 million in the year 2016, such eye tracking devices can give them a feel of self reliability \cite{census2016report}.    

Another alternative to using physiological signals for the control of wheelchair is to track the eye movement of the patient. The history of eye tracking dates back to mid 19\textsuperscript{th} century. Some of the earlier applications of eye tracking were in text detection, text reading, to measure the effectiveness of advertisements, to control devices using the eye \cite{advert2012}. Nowadays, eye tracking is used in scientific research, augmented reality, medical field and smart devices\cite{augmented2016,medical2017,smartDevices2014}. The system requires simple cameras clubbed with image processing and pattern recognition algorithms and can be implemented in commonly available processors. This also brings down the cost of the overall control system in comparison to those which uses physiological signals\cite{ecg2018,eeg2013}.  
This project aims to make a reliable eye tracking model and implement the hardware. This model should be able to accurately and efficiently track the eyeball movement and assign commands to each unique movement, to move a bot/ wheelchair \cite{gajwani2010eye,gautam2014eye,plesnick2014eye}.

\begin{figure}[h]
\centering
\includegraphics[width=7cm, height=10cm]{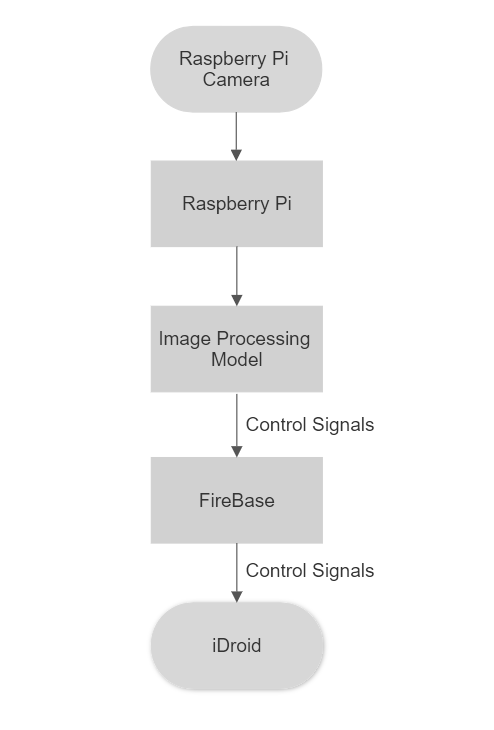}
\caption{Flowchart}
\label{fig:FlowChartOverall}
\end{figure}

 This model can be helpful in different fields.\newline
The following flowchart shown in Fig. \ref{fig:FlowChartOverall} gives our approach on the implementation of the model.


\section{Eye Tracking}
say generally about the eye tracking algorithm with the overall flow chart here.
\subsection{Hardware Setup}
\begin{itemize}
\item Raspberry Pi Camera: It is a 5 megapixel camera module which was set to record video at 60 frames per second (fps).It  has  been  highly  optimized  to  work  with  the  Raspberry  Pi Processor. It has extremely fast transmission speeds due to its Parallel-In-Parallel-Out (PIPO) set of ribbon cables. The RPi camera  is  mounted,  at  a  distance  of  9cm  to  12cm,  in  front of  the  eye  for  the  best  focus.  This  RPi  camera  then  captures the  movements  of  the  pupil  of  the  eye  and  sends  it  to  the RaspberryPi for processing.
\item RaspberryPi: It is  a  very  powerful  and  low-cost  micro-processor.  It  has  a  1GB  RAM  capacity  and  supports  many of  the  common  python  libraries.  The  RaspberryPi  is  used  to process the incoming video frames from the RPi Camera. The incoming  frame  from  the  RPi  camera  is  a  three-dimension colour image This coloured image was first made into a greyscaled image and then resized to a 128px by 128px image.This  was  done  to  reduce  the  overall  computational  time.Then this resized gray scaled image is passed onto the image processing model sitting on the RP

\end{itemize}

\subsection{Image Processing and control command generation}

The image processing model we used is a convolutional neural network trained on a self created dataset. It receives the image from rpi local host, processes it and predicts the orientation of the eye. The predicted orientation is passed on to the firebase.\\
The dataset is self made using pictures clicked of around 10 persons. The dataset distribution is as follows:\\
Down: 2045 images\\
Up: 2475 images\\
Left: 2858 images\\
Right: 2797 images\\
Straight: 3058 images\\
The images are of various sizes in the RGB(color) format. 
\begin{figure}[h]
\centering
\includegraphics[width=8cm, height=13cm]{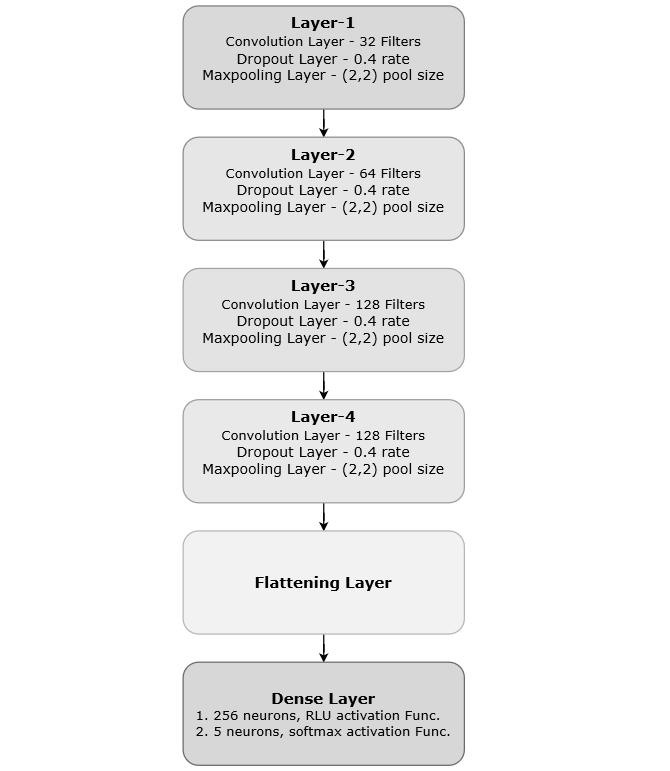}
\caption{CNN Flowchart}
\end{figure}

\subsubsection{Data Preprocessing}
In the data preprocessing phase, all the images are cropped to focus only on the eye and the images are converted to gray scale. The images are further resized to 128 by 128 pixels to have a uniform size throughout the dataset. 

\subsubsection{Neural Network architecture}

As the flow chart depicts, the neural network consists of four sets of convolutional layer, dropout layer and max pooling layer followed by flattening layer and full connection using dense layer. The convolutional layers have 32, 64, 128 and 128 filters in successive layers. The kernel size of these filters are fixed to 3 by 3. Dropout layers are added to reduce over-fitting with a dropout rate of 0.4. The kernel size in max pooling layers are fixed to 2 by 2. A flatten layer flattens out all the inputs received by it which is then fed to fully connected dense layer with 256 neurons. The output of this layer is given to the final output layer with 5 neurons which predict the orientation. The convolutional layers and deep dense layer uses rectified linear units as activation function. The output layer uses softmax to predict the probabilities that the input belongs to a certain class.
\begin{figure}[h]
\centering
\includegraphics[width=8cm, height=6cm]{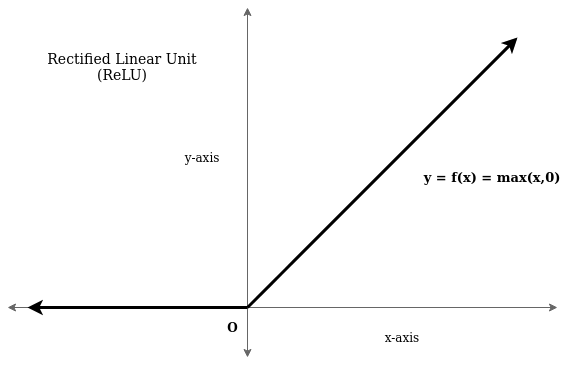}
\caption{Rectified Linear Unit}
\end{figure}
\begin{figure}[h]
\centering
\includegraphics[width=8cm, height=6cm]{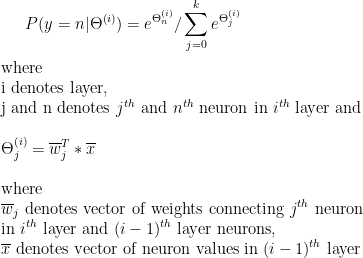}
\caption{Softmax Function}
\end{figure}

\subsubsection{ Training Phase}

In the training phase, the processed images are first loaded. The entire dataset is further divided into training and validation sets randomly. The validation set is approximately 30\% of the total dataset. The model is trained with adam optimizer and categorical cross-entropy as loss function for 25 epochs. The trained model is saved.

\section{Mobile Application}
An Android Application is made to enable manual override of the bot during certain situations. It has options to update the direction and even the speed of the bot to the firebase. It also allows the user to give voice commands. The android application acts as a safety precaution in case the RaspberryPi is unable to send controls and communicate with firebase.
\section{iDroid}
iDroid is the hardware prototype of a wheelchair. The prototype of the wheelchair is composed mainly Arduino UNO, NodeMCU, Ultrasonic Sensor (HC-SR04) and L293D Motor Driver Shield. The Arduino and Motor Driver Shield are interfaced and made into a single unit.  The Motors and the Ultrasonic Sensor are interfaced to the Motor Driver Shield. A power bank with dual port is used to power the Arduino and the Motor Driver Shield for better performance.
The firebase has a tag named Signals which is updated according to the eyeball movement of the user. The following table indicates the relation between the movement of the eye and the movement of the bot. These signals will be considered valid and get updated to the firebase only if the same signal is given by the user for more than 30 frames.
This information is then received from the firebase using the NodeMCU. This received information is then communicated to the Arduino using the SoftwareSerial library. The Arduino now receives the information and signals the Motor Driver Shield to appropriately move the bot. All this happens with a delay of about 500ms. When the bot is in motion the Ultrasonic Sensor continuously checks for any obstacles. If any obstacle is detected within the threshold distance, the bot stops immediately.
\includegraphics[width=8cm, height=6cm]{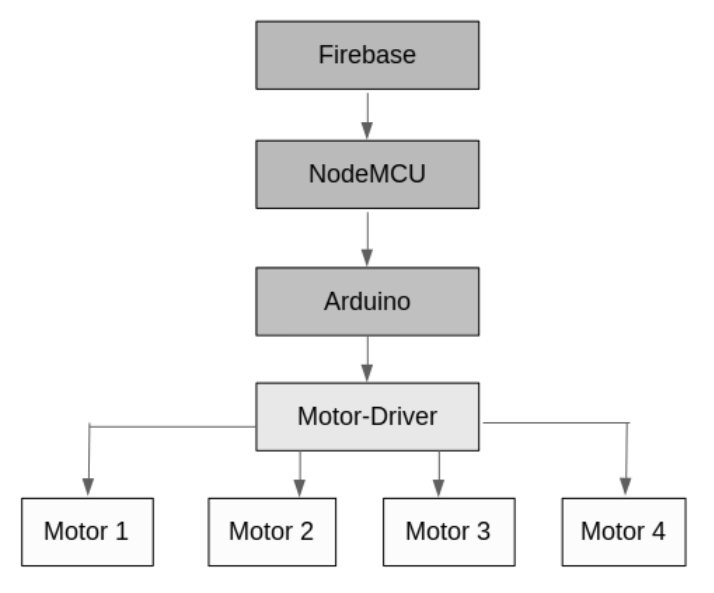}
\subsection{NodeMCU}
NodeMCU was chosen for this project because of its ability to connect easily to a nearby Wi-Fi Router or a Mobile Hotspot. It is small in size, inexpensive and has low power requirements. It also has a plenty of online learning resources and help from the online community. 

Moreover it can be programmed using the Arduino IDE by installing some necessary board packages and thereby reduces the effort of getting used to a new programming environment. The GPIO pins are used for Serial Communication with the help of the SoftwareSerial Library.

The NodeMCU connects to a nearby Wi-Fi Router or a Mobile Hotspot and fetches the information from the Firebase. After this it communicates the information to the Arduino using the SoftwareSerial library.
\begin{table}[htpb]
\centerline{
\begin{tabular}{|c|c|}\hline

Movement of the eye&Movement of the bot\\\hline
Down&Stop\\\hline
Left&Left\\\hline
Right&Right\\\hline
Up&Start\\\hline
Straight&Forward\\\hline
\end{tabular}}
\caption{Control Signal}
\end{table}
\subsection{Arduino UNO}
Arduino was chosen for this project because it can easily connect with different sensors and has got many shields which provide it with additional capabilities. It also has got a large support from the Arduino community which is very active. Arduino UNO was chosen because it is inexpensive and small in size. It also is the most popular and the most used board among the other microcontrollers of the Arduino Family and a result has very good documentation too.

The Arduino is programmed using the Arduino IDE. The TX and RX pins on the board are the pins meant for Serial Communication. The other pins can be used for Serial Communication by using the SoftwareSerial Library.

The information that the NodeMCU receives from the Firebase is communicated to the Arduino using the SoftwareSerial library. The Arduino receives the information and signals the Motor Driver Shield to appropriately the move to the bot according to the command received.
\subsection{L293D Motor Driver Shield}
L293D Motor Driver Shield was chosen because of its ability to power 4 bi-directional DC Motors. It also allows easy interface of additional sensors thereby eliminating the need of a breadboard.

The Motor Driver Shield offers an 8 bit speed control (0-255) for varying the speed of the bot. The speed of the bot is gradually increased to the required value to prevent sudden jerks to the user and also sudden loading of the power source. The use of the AFMotor Library simplifies the coding part as it has direct inbuilt functions for controlling the motors.
\subsection{Ultrasonic Sensor (HC-SR04)}
Ultrasonic Sensor was chosen for this project because it is inexpensive, easy to use, small in size, available readily, highly accurate and detects most type of obstacles. When the bot is in motion the Ultrasonic Sensor continuously checks for any obstacles. If any obstacle is detected within the threshold distance, the bot stops immediately.\\
\begin{figure}[h]
\centering
\includegraphics[width=8cm, height=6cm]{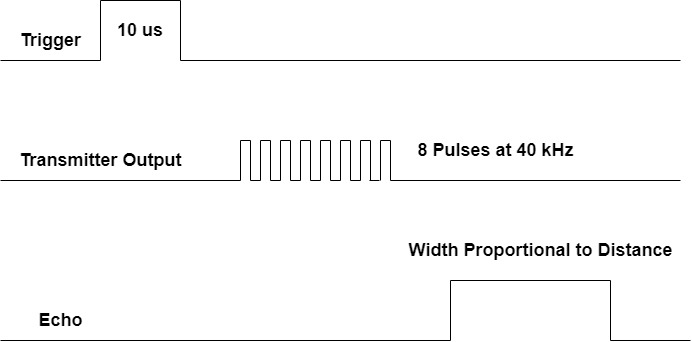}
\caption{Ultrasonic Timing Diagram}
\end{figure}
The Ultrasonic Sonic Sensor consists of 2 Ultrasonic Transducers which act as Transmitter and Receiver. The Trigger Pin of the Ultrasonic Sensor is given a pulse of a minimum duration of 10 µS, resulting in the transmission a sonic burst of 8 pulses with a frequency of 40 KHz. Once this is transmitted the Echo pin goes high and then goes low only if the pulses reflect from the surface of an obstacle and reach the sensor. If there are no obstacles encountered then the echo signal goes low by itself after 38 mS.

Hence once we know the time and speed, we estimate the distance of the obstacle from the bot and its accuracy is further increased by using the NewPing library.

\section{Results and Discussions}
\subsection{Performance of the eye tracking algorithm}

The confusion matrix is shown in Tab
\begin{table}[htpb]
\centerline{
\begin{tabular}{|c|c|c|c|c|c|}\hline
Actual&\multicolumn{5}{|c|}{Predicted class}\\\cline{2-6}
class&Down&Left&Right&Up&Straight\\\hline
Down&628&0&0&0&0\\\hline
Left&0&851&0&0&0\\\hline
Right&0&1&840&0&0\\\hline
Up&0&1&0&714&3\\\hline
Straight&0&1&0&0&931\\\hline
\end{tabular}}
\caption{Confusion Matrix.\label{tab:ConfMat}}
\end{table}

The classification report is as follows:
\begin{table}[htpb]
\centerline{
\begin{tabular}{|c|c|c|c|c|}\hline
Classes&Precision&Recall&F1 Score & Support\\\hline
Down&1&1&1&628\\\hline
Left&1&1&1&851\\\hline
Right&1&1&1&841\\\hline
Up&1&0.99&1&718\\\hline
Straight&1&1&1&932\\\hline
\end{tabular}}
\caption{Classification Report\label{tab:ClassReport}}
\end{table}

\subsection{Overall system performance}
\section{Conclusions}

\begin{itemize}
  \item 
   The RPi camera needs sufficient illumination to capture the images correctly. The prediction on the RPi camera is slow as it requires high computation power to process the convolutional neural network. Also, RPi faced compatibility issues with TensorFlow versions. So RPi just resizes the captured images, grayscale’s the image and feeds this to the computer to do the predictions. The resizing of images and conversion from color to grayscale is done to reduce latency issues. It should be ensured that the RPi camera focuses only on the eye and doesn’t move much.
   \item The predictions are fast enough to detect even the quick blinks. To avoid the detection of unintentional blinks, the firebase is not updated until the predictions are the same for a minimum of 20 frames. This also reduces latency issues in firebase.
  \item The four-wheeler iDroid being used requires a lot of power. So the iDroid must be powered by well charged Power Bank.
  \item \textbf{\large{}}
  The training accuracy is found to be 99.99848\%. This indicates that there is a chance that the model is overfitting. Attempts has been made to make the model more robust and error-free by adding dropout layer and making the data set skew free.   
    \item \textbf{\large{}}
  The image processing model(CNN) is able to process and predict at a rate of 15-16 FPS.
\end{itemize}

\bibliographystyle{IEEEtran}
\bibliography{Main_Conf}	
\end{document}